\documentclass[aps,nofootinbib,showpacs,twocolumn,longbibliography]{revtex4-1}

\usepackage[CJKbookmarks, pdftex, bookmarksnumbered, bookmarksopen, colorlinks, citecolor=blue, linkcolor=blue]{hyperref}

\usepackage{amstext,amsmath,amssymb,amsfonts,bbm}
\usepackage[latin1]{inputenc}
\usepackage{graphicx}
\usepackage{epstopdf}
\usepackage{amsthm}
\usepackage{tocvsec2}
\usepackage{enumerate}
\usepackage{subfigure}
\usepackage{color}

\newcommand{\vS}{\widetilde{\mbox{\boldmath$\epsilon$}}}

\usepackage{multirow} \usepackage{rotating}

\def\beq{\begin{equation}}
\def\be{\begin{equation}}
\def\ee{\end{equation}}
\def\bes{\begin{eqnarray}}
\def\ees{\end{eqnarray}}

\def\tr{{\rm tr}}

\begin{document}

\title{Spacetime thermodynamics without hidden degrees of freedom}

\author{Goffredo Chirco, Hal~M.~Haggard, Aldo Riello, Carlo Rovelli}

\affiliation{Aix Marseille Universit\'e, CNRS, CPT, UMR 7332, 13288 Marseille, France.\\
Universit\'e de Toulon, CNRS, CPT, UMR 7332, 83957 La Garde, France.}

\date{\today}

\begin{abstract} 

\noindent 
A celebrated result by Jacobson is the derivation of Einstein's equations from Unruh's temperature, the Bekenstein-Hawking entropy and the Clausius relation. This has been repeatedly taken as evidence for an interpretation of Einstein's equations as equations of state for unknown  degrees of freedom underlying the metric.  We show that a different interpretation of Jacobson result is possible, which does not imply the existence of  additional degrees of freedom, and follows only from the quantum properties of gravity.  We introduce the notion of quantum gravitational Hadamard states, which give rise to the full local thermodynamics of gravity. 
\end{abstract}
\maketitle

\section{Introduction}

In a celebrated paper \cite{Jacobson:1995ab}, Ted Jacobson presented a surprising thermodynamical derivation of Einstein's equations, based on three inputs: (i) the vacuum of a quantum field theory in Minkowski space behaves as a thermal state at the Unruh temperature
\be T=\frac{\hbar a}{2\pi}
\ee
for an  observer with acceleration $a$; (ii) there is a \emph{universal} entropy density $\alpha=1/4\hbar  G\label{bekensteine}$ per unit area, associated to any causal horizon in a locally Minkowski patch of spacetime, giving an entropy
\be 
S= \frac{A}{4\hbar G}\label{unruht}
\ee 
for a horizon region of area $A$;  and  (iii) a local entropy balance relation 
\be 
\delta S= \frac{\delta E}{T}
\label{clausiusr}
\ee 
holds, where $\delta E$ is an energy exchange.  By interpreting $\delta E$ as the energy of matter flowing across the local Rindler horizon of the accelerated observer, and matching the variation of the area with the focusing effect of spacetime curvature, Jacobson was able to show that if the three equations above hold for \emph{any}  \emph{local} frame, then Einstein's equations follow. 

This is a beautiful piece of theoretical physics. But its interpretation is not clear.  A common understanding \cite{Jacobson:1995ab,Padmanabhan2009,Verlinde2011a} takes the result as evidence that Einstein's equations have a statistical origin and should be interpreted as equations of state for unknown underlying  degrees of freedom, with the metric being a macroscopic ``coarse-grained" variable.  In this paper we show that a different interpretation is possible. 

The alternative interpretation, which we develop mostly following \cite{Bianchi:2012br,Bianchi:2012ev}, is based on the fact that the gravitational field has quantum properties. The microscopic degrees of freedom are those of the quantum gravitational field and the Einstein equations express only the classical limit of the dynamics. The entropy across the horizon measures the entanglement between adjacent spacetime regions.  Its finiteness is evidence for the quantization of the gravitational field: this is analogous to the fact  that the finiteness of the black-body electromagnetic entropy is evidence for the quantization of the electromagnetic field. 

We show that the Jacobson result is consistent with this simpler and tighter scenario.  The finiteness and the universality of the entanglement entropy across spacetime regions indicates ultraviolet quantum discreteness, as it did for Planck and Einstein at the beginning of the XX century. 

Section \ref{sec:entropy} summarizes basic facts about entropy in quantum theory. Section \ref{sec:entanglement} discusses the relation between entanglement entropy and temperature.   In Section \ref{sec:Jacobson} we interpret Jacobson's result in these terms.  In Section \ref{eu} we discuss the dependence of the entropy on the number of fields. 

In Section \ref{sec:quantumgravity} we outline how the scenario we consider can be concretely implemented in loop quantum gravity. In this context we introduce the Hadamard quantum gravitational states. In Section \ref{sec:architecture} we discuss the meaning of these states, which give rise to the correct entropy and  provide a foundation for the semiclassical states of the geometry, in the sense of  \cite{Bianchi:2012ev}. We summarise our results in the last Section. In the three appendices, we discuss the relation between thermometers and KMS states (Appendix \ref{A}), as well as the relation between the simplicity conditions and the area-energy relation (Appendix \ref{B}); finally, in Appendix \ref{C}, we recall the main lines of Jacobson's original argument.  

\section{Statistical and Entanglement entropy}\label{sec:entropy}

We start by recalling some basic facts about entropy in quantum systems. Consider a quantum system whose pure states are described by a normalised vector $\psi$ in a Hilbert space $\cal H$.  A generic, not necessarily pure, state is described by a density matrix $\rho:{\cal H}\to{\cal H}$ satisfying $\tr[\rho]=1$ and $\rho^\dagger=\rho$. To any such density matrix we can associate the von Neumann entropy
\be
  S_{\rm vN}=-\tr[\rho\ln\rho].
\ee
This entropy measures the lack of information about measurement outcomes that is there \emph{in addition} to that implied by the Heisenberg uncertainty relations.  It is important to distinguish two extreme examples of sources for such uncertainty.  The difference is at the core of the argument of this paper:\\

\paragraph{Statistical entropy.} Consider a system with many degrees of freedom and a small number of macroscopic observables describing it. Fix a value $a_i$ for these observables. The density matrix $\rho_{a_i}$ representing this information about the system is the projector onto the eigenspace determined by these values.  The dimension $n$ of this Hilbert space counts the number of (orthogonal) states compatible with these values.   The microcanonical von Neuman entropy of $\rho_{a_i}$
\be
    S_{\rm vN}= -  \sum_{i}  p_i \ln p_i = \ln n 
\ee
measures our ignorance of the microstate, where the probability of every microstate is $p_i=1/n$. This is the standard entropy of statistical mechanics. (In a similar manner one can construct the canonical version of the density matrix.)\\
\paragraph{Entanglement entropy.}  Consider now one of the subsystems of a larger system formed by two parts, described by the Hilbert spaces $\cal H$ and  $\cal H'$, respectively. A generic pure state of the composite system can be written in the form 

\be
        |\Psi\rangle=\sum_{n, m'}c_{n m'}\, |n\rangle\otimes |m'\rangle
\ee
where  $|n\rangle$ and  $|m'\rangle$ are bases in the respective spaces. If we consider only measurements performed on the first system, these are fully characterised by the density matrix 
\be
       \rho= \tr_{\cal H'} |\Psi\rangle\langle\Psi |=\sum_{n} p_n  |n\rangle \langle n|
\ee
where 
\be
p_n = \sum_{m'}  |c_{nm'}|^2. 
\ee
The von Neumann entropy is  
\be
 S_{\rm vN}= -  \sum_{n}  p_n \ln p_n 
\ee
and measures the amount of entanglement between the two systems. 

Therefore we see that the von Neumann entropy can describe two (related but) distinct physical quantities: in the first example, it describes our ignorance of the specific microstate; in the second, it is a measure of entanglement. The first describes simply a translation of the  standard classical entropy of Gibbs and Boltzmann. The second is a genuine quantum phenomenon, which does not exist in classical physics. 

\section{Entanglement thermodynamics}\label{sec:entanglement}

When the von Neumann entropy describes statistical ignorance, its relation with thermodynamical entropy is manifest.  In fact, the thermodynamics of quantum systems is described in all textbooks in terms of von Neumann entropy.   Is there a relation between, thermal quantities and von Neumann entropy when it describes entanglement?  

Entanglement entropy is different from standard statistical entropy in some respects, but several characteristic thermodynamical relations remain valid \cite{Deutsch2013}.  Let us illustrate this fact in a simple case.  Consider a system with two degrees of freedom, respectively described by the Hilbert spaces $\cal H$ and  $\tilde{\cal H}$. In the tensor product Hilbert space consider the \emph{pure} state 
\be
        |\Psi_\beta\rangle = \sum_n e^{-\frac{\beta}2 E_n}|n\rangle\otimes |\tilde{n}\rangle.
        \label{states}
\ee
where $\beta$ is a parameter that we keep free,  $H|n\rangle=E_n|n\rangle$ in  $\cal H$, where $H$ is the hamiltonian, and the basis $|\tilde{n}\rangle$ in  $\tilde{\cal H}$ is an arbitrary orthonormal basis.  This is a highly entangled pure state. The corresponding density matrix obtained tracing over  $\tilde{\cal H}$ is 
\be
        \rho_\beta= N_\beta\, e^{-\beta H} \label{rr}
\ee
where $N_\beta$ is the normalization factor needed to have $\tr [\rho]=1$. The expectation value of the energy $H$ in the reduced state is 
\be
        E(\beta)=\langle \Psi_\beta |H| \Psi_\beta \rangle= \tr[\rho_{\beta} H]=\langle H\rangle_{\rho_{\beta}} \label{energy}
\ee
and the von Neumann entropy is given by
\be
S(\beta)=-\tr[\rho_\beta \ln \rho_\beta]. 
\ee
Restricted to $\cal H$, the state looks ``thermal'' even if the systems as a whole is in a pure state and does not have many degrees of freedom. Now, consider a small variation $\delta|\Psi_{\beta} \rangle$ of the state $|\Psi_{\beta} \rangle$. The change in entanglement entropy $S(\beta)$ is proportional to the change in the averaged energy, namely
\begin{eqnarray}
\delta S&=&-\tr[\delta \rho_\beta \ln \rho_\beta]
= \beta\, \tr[\delta \rho_\beta H ] = \beta\, \delta \langle H\rangle. 
\label{de S}
\end{eqnarray}
where the first equality follows from the identity $\tr[\delta \rho_{\beta}]=0$ and the second from the Gibbs form of \eqref{rr}. 

By using \eqref{energy} and defining
\be
T \equiv \frac1{\beta}, \label{t}
\ee 
equation \eqref{de S} takes the form of a thermodynamical relation
\be
\delta S= \delta E/T. \label{clau}
\ee
Equation \eqref{clau} is a relation of equilibrium. However, there is no thermodynamical limit, no large number of degrees of freedom in the argument above. In the simple context of a two-degrees of freedom system, $T$ does not have an interpretation as temperature.  

On the other hand, if we have many copies of this coupled system, and a thermometer coupled with them all, then $T$ can be interpreted as a temperature. Indeed it determines the raising and lowering transition probabilities between thermometer eigenstates that thermalizes to the probability distribution 
\be
     p\sim e^{-\beta E}
\ee
where $E$ is the energy difference between the thermometer states.  This result follows from the fact that \eqref{rr} is a KMS state for the time flow, and its proof is recalled in Appendix  \ref{A}. 
 
The example above might sound artificial, since the one-parameter family of states \eqref{states} was introduced ad hoc.  But it is not so artificial: in fact, its structure exemplifies that of the Minkowski vacuum $|0_M\rangle$ under the split of the Fock space into the tensor product of a Hilbert space capturing the degrees of freedom accessible from within the horizon of a uniformly accelerating observer, and those outside (Rindler wedge localization) \cite{Haag:1992hx}.  The Minkowski state is a pure state and when restricted to the Rindler wedge observables it turns out to be given by the density matrix 
\be
     \rho = \tr_{\tilde{\cal H}}|0_M\rangle\langle 0_M|= N e^{-\frac{2\pi}\hbar K}
     \label{rhok}
\ee 
where $K$ is the boost generator. This is the content of the Bisognano-Wichmann theorem \cite{Bisognano:1976za}, which can be derived rigorously from quantum field theory axioms, or  formally with a variety of simple manipulations.  It is a consequence of Lorentz invariance and positivity of the energy. 

The motion of a uniformly accelerated observer is precisely generated by $K$. The proper time $\tau$ of an observer moving at constant acceleration $a$ is given by $\tau=\eta/a$, where $\eta$ is the boost parameter of the transformation generated by $K$. Therefore once rescaled by $a$, $K$ generates proper time translations on the world line of a uniformly accelerated observer, that is 
\be
      H=a K \label{aK}
\ee
is the generator of the evolution in the proper time in the Rindler wedge. As a consequence, the density matrix that represents the outcome of measurements on the spacetime region bounded by the horizon of an accelerated observer in the \emph{pure} state $|0_M\rangle$  can be rewritten as 
\be
    \rho= \tr_{\tilde{\cal H}}|0_M\rangle\langle 0_M|= N\, e^{-\beta H},
     \label{rhoa}
\ee 
where 
\be
   T=\frac1\beta= \frac{a\hbar }{2\pi},
   \label{unruh}
\ee
which is the Unruh temperature associated to an observer accelerating in the Minkowski quantum-field vacuum \cite{Unruh:1976db}. The Unruh temperature is a physical temperature: it determines the transition probabilities of a system interacting with the quantum field and moving along the accelerated trajectory. 

This shows that under appropriate conditions the quantum mechanical entanglement can define a temperature which is the temperature measured by a physical thermometer. In particular, this happens when the density matrix obtained by reducing the state to a subalgebra has the form  \eqref{rr} for a hamiltonian $H$ generating a flow that can be identified with the time evolution of the thermometer. 

Parenthetically we note that the density matrix associated to a subsystem  can \emph{generically} be written in the form 
\be
   \rho\sim e^{- H}
\ee
for \emph{some} hermitian operator $H$. It suffices to define $H$ as minus the log of $\rho$.  The quantity $H$ defined in this manner is called the ``entanglement hamiltonian'', ``modular hamiltonian''\cite{Haag:1992hx} or ``thermal time'' \cite{Rovelli:1993ys,Connes:1994hv} hamiltonian in various contexts. It generates an ``evolution'' in the Hilbert space $\cal H$, and the state $\rho$ is a KMS state with respect to this flow. In general this flow does not correspond to a flow in spacetime.  In the case of the restriction of vacuum to a Rindler wedge, it does. 
   
The thermodynamical aspects of the setting considered by Jacobson can be understood in this framework. There is a quantum field in its vacuum seen by an accelerated observer. The observer interacts only with a sub algebra of the field observable algebra, and therefore describes the state of the field in terms of a density matrix \eqref{rhoa}. The entropy that quantifies the incertitude, due to entanglement, of an observer's measurement, and the expectation value of the generator \eqref{aK} of his proper-time translations are related by \eqref{de S}, namely
\be
\delta S=\delta E/T. 
\ee
The Unruh temperature and the entropy
\be
   \delta S= \frac{2\pi }{a\hbar}\,  \delta E
 \label{dS}
\ee
can be computed directly from the form of the entangled state $\rho$. These are relations that pertain to standard quantum field theory and do not require any additional hypothesis about underlying degrees of freedom to be true. They depend on the local Lorentz invariance of the vacuum and in particular on its local (short scale) consequences.  General Relativity plays no role so far. Equation \eqref{dS} has nothing to do with the Einstein equations.  In particular, the relation $\delta S=\delta E/T$ knows nothing about the Newton constant or the Einstein equations.  How do then Einstein's equations emerge in Jacobson's argument?

\section{The Jacobson result}\label{sec:Jacobson}

The key ingredient of Jacobson's derivation is the assumption that the entropy density per unit horizon area is not only finite, but, most importantly, \emph{universal}.  That is, in the relation 
\be
S=\alpha A, \label{key}
\ee
it is assumed that $\alpha$ is finite and does not depend on any detail of the physical situation. Here, $A$ is the area of the 2-dimensional spacelike bifurcation surface of the causal horizon comprising the wedge.

This universality is a strong assumption that falsely may appear to be innocent. The fact that the entropy across a region is proportional to the area is common when correlations are sufficiently short scale compared to the geometry of the surface.  However, in general the proportionality constant will depend on the system and on the state. Jacobson assumes it does not. 

The entropy density of the density matrix \eqref{rhok} is infinite, because of the contribution of arbitrarily short wavelength correlations across the Rindler horizon. If we cut off the theory at some length $\ell_p$, we obtain an entropy that scales as $1/{{\ell}_p}^2$ \cite{Sorkin1983,Bombelli1986, Srdnicki1993, Solodukhin2011}, that is
\be
    S=\frac{\alpha_c}{\ell_p^2} A
\ee
or $\alpha=\frac{\alpha_c}{\ell_p^2}$.  If we change the state, the area does not change (we are on a fixed geometry) but the entropy does change, therefore 
\be
\delta S=\delta(\alpha A) = (\delta \alpha)\,A. 
\label{dkey}
\ee
If one instead assumes, as Jacobson does, that the proportionality constant between entropy and area is universal, then one either gets a contradiction (because one can change the entropy by changing the state), or one has to relax the fact that the geometry is fixed and allow it to change with a change in the state, thus allowing the area to vary
\be
\delta S = \alpha \,\delta A.
\label{ddkey}
\ee
As we have seen above the entanglement entropy across a surface satisfies the relation $\delta S =\delta E/T$, where $E$ is an energy. Therefore we obtain 
\be
\delta E = T \alpha \,\delta A, 
\label{ddkey}
\ee
and as a consequence if the constant $\alpha$ is to be universal, area and energy variations must be related. Inserting the value of $T$ from equation \eqref{dS} into \eqref{ddkey} leads to the fundamental relation
\be
 \delta E= \frac{a\hbar}{2\pi}  \alpha \,\delta A. 
\label{ar_en}
\ee
Notice that this relation has nothing thermodynamical in it.  It is a relation between energy and geometry.  
This relation was derived by Frodden, Gosh and Perez \cite{Frodden:2011eb} in a different context as a consequence of the Einstein equations (see also \cite{wald}).  The Frodden-Gosh-Perez relation reads 
\be
\delta  E  =\frac{a}{8\pi G}\, \delta  A  \label{fgp}.
\ee
and agrees with \eqref{ar_en} identifying the universal constant $\alpha$ as $\alpha=1/4\hbar G$. 

Thus, what Jacobson has shown in his derivation (which we recall in Appendix \ref{C}) is that equation \eqref{fgp} is not only a consequence of Einstein's equations, but is also a \emph{sufficient} condition for the Einstein equations to hold. If we assume the validity of \eqref{fgp} in any frame, then the Einstein equations follow. In this sense, Einstein's equations are encoded in the proportionality between classical variation of energy and horizon area, as measured by a uniformly accelerating observer. 

By itself, the fact that the Einstein equations can be encoded in a single simple equation relating an energy and a geometrical quantity is well known. Of course the trick is requiring that this holds everywhere and in any frame, thus adding general covariance as an independent input. For instance, see the nice paper by John Baez and Emory Bunn \cite{Baez:2001qy} where it is  pointed out that Einstein's equations follow entirely from the request that the second variation of the volume of a small sphere of particles is proportional to the sum of the energy and momentum flow components in the centre of the sphere. Jacobson's is an elegant and null version of Baez and Bunn's timelike result. 
 
The important point of this discussion is that Jacobson's result can be split into two parts. The second part being the step from \eqref{ar_en} to the Einstein equations.  This is a nice piece of differential geometry, but has nothing to do with thermodynamics.  The first step is to get to equation \eqref{ar_en}. As we have seen, this equation comes from considering the entanglement entropy across the horizon plus a universality assumption. Again, here statistical considerations play no role, and nothing points to underlying degrees of freedom, or to a reading of the Einstein equations as equations of state. The key is simply the universality of the entanglement entropy-density. We comment on this below. 

\section{Universality:  the species problem} \label{eu}

The interpretation of the Bekenstein entropy as entanglement entropy has been called into question precisely due to the unclear dependence of the entanglement entropy on the number of species at the cutoff scale. How can entanglement entropy not be dependent on the number of species? 

A recent result \cite{Bianchi:2012br} has shown that for a low energy perturbation of the Minkowski vacuum, the variation of entanglement entropy $\delta S$  across the Rindler horizon is independent from the number of matter species and independent from the ultraviolet cut-off, provided one includes gravitons together with matter fields, in a perturbative approach. What happens is that the change in the energy of the vacuum $ \delta E$, expressed in terms of the energy-momentum of matter \emph{and} gravitons $T_{\mu \nu}$, is related to the d'Alambertian of the graviton field by means of the perturbative Einstein equations,
\be
\Box h_{\mu \nu}= -\sqrt{8\pi G} (T_{\mu \nu}-\frac1 2 \eta_{\mu \nu} T^{\sigma}_{\sigma}), 
\label{epertu}
\ee
from which 
\be
\delta E  =\frac{a}{8\pi G}\, \delta A \label{reja}
\ee
follows. The change in the area of the causal horizon has now a \emph{semiclassical} interpretation. Classical background geometry is not affected, and the perturbation of the area is determined by the quantum gravitational field $h_{\mu \nu}$ and given by the expectation value of the operator $\hat{A}$ associated to the first order \textit{variation} of the area with respect to its background value, as a function of $h_{\mu \nu}$. That is $\delta A= \delta \langle \hat{A} \rangle\equiv \tr[\delta \rho \hat{A}]$. 

This perturbation controls the deflection of light rays related to the expansion of the area of the horizon surface, so that equations \eqref{clau}, \eqref{unruh} and \eqref{reja} together provide a dynamical Bekenstein-Hawking area law relation, 
\be
\delta S=\frac{\delta A}{4G\hbar}, \label{beke}
\ee
which relies on the entanglement of the gravitationally dressed vacuum fields.

Importantly, equation \eqref{beke} does not depend on the physics at the UV scale, as $\delta S = \delta E/T$ involves only differences in the low energy modes, while its universal character has a dynamical origin: it derives from the universal coupling of gravitons to the energy momentum tensor.

At the perturbative level, this result provides a connection between Bekenstein-Hawking entropy and entanglement entropy and indicates that the relation between area and energy variations in \eqref{reja} can only have a dynamical origin: it requires the Einstein equations. 

This connection between Bekenstein-Hawking entropy and entanglement entropy is not in terms of statistical ignorance about mysterious underlying microscopic degrees of freedom, it does not imply any ``thermal'' characterization of the Einstein equations.

The universality of $\alpha$ is the effect of an active role of gravitational dynamics at the quantum level. One can gain an intuitive understanding of the fact that gravity cuts the degrees of freedom off in a universal manner as follows. The quantum gravitational regime starts when the total energy density reaches the Planck scale.  Matter fluctuations interact gravitationally, so in the presence of more than one field, say $n$ fields, this regime is reached at a larger length scale than in the presence of a single field.  More precisely, if we introduce a cutoff at the scale $\ell$, on dimensional grounds the vacuum energy density of the fluctuations of \emph{a single} field is proportional to $\ell^{-4}$. Within an $\ell^3$ volume, the (negative) gravitational potential energy of these fluctuations goes as $G \ell^{-3}$ and is proportional to $n^2$ because each field interacts with any other field. We enter in a strong gravitational field regime at the scale where these two energies balance, namely at the length scale $\sqrt{n\hbar G}$.  Thus, for a single field we expect a gravitational cutoff at the Planck scale $\sqrt{\hbar G}$, but for $n$ fields the cutoff scale is reached at a larger scale by a factor $\sqrt{n}$, that is, the cutoff scale $\ell$ depends on the number of fields as
\be
   \ell\sim \sqrt{n}.
\ee
The entanglement entropy of $n$ fields is 
\be
       S\sim n \frac{A}{\ell^2}
\ee
where $\ell$ is the cutoff scale.  The last two equations together indicate that $S$ does not depend on the number of fields \cite{Bianchi}. It is the universality of gravity, not statistics, that determines the entanglement entropy universality of the entropy density. 

So far, we have stayed away from the Planck scale. It is time to ask what may happen there and complete the picture. 

\section{Finiteness: the Jacobson result from quantum gravity}\label{sec:quantumgravity}

At the beginning of the XX century, Planck and Einstein were led to the existence of a fundamental discreteness in the structure of the electromagnetic field, and in fact to the discovery of the existence of the photon,  from the absence of the ``ultraviolet catastrophe," namely from the finiteness of the entropy of the electromagnetic field.  The entropy of black body radiation, indeed, is given by 
\be
    S= \frac43\ \sqrt[4]{\frac{\pi^2VU^3}{15 c^3 \hbar^3}},
\ee
where $V$ is the volume of the box and $U$ the energy, and this diverges if we take $\hbar\to0$.  Quantum theory is the source of the Planck-size granularity that renders phase space volumes finite, and therefore avoids the divergence of the entropy.  In the entropy \eqref{beke}, associated to a gravitational field, $\hbar$ is in the denominator as well. This indicates that it is quantum mechanics that cuts off the divergence of the entropy. In fact, the formula indicates that there should be a physical (real) quantum mechanical cutoff at the Planck scale.

This is precisely the conclusion that one obtains from a full quantum gravitational treatment of the problem in the context of loop quantum gravity,
where the \emph{discreteness} of the geometry at the Planck scale appears as a conventional quantization effect: the gravitational field determines lengths, areas and volumes; since the gravitational field is a quantum operator, these quantities are given by quantum operators. Planck scale discreteness follows from the spectral analysis of such operators, which indicates minimum non-vanishing sizes at the Planck scale. This is the key result of the loop theory and is responsible for the UV finiteness of the entropy density. 

Let us see how Jacobson's result emerges  from this framework.  For this, we recall some basic notions from the full theory, relevant to the present discussion \cite{Rovelli:2011eq,Rovelli}.

In loop gravity, quantum states of the geometry are described by $SU(2)$ spin networks.  Let us consider here for simplicity a single link of the spin network.  The corresponding quantum state is a function $\psi(U)$ on $SU(2)$, and $U$ is classically interpreted as the open path holonomy of the gravitational Ashtekar connection along the link. A basis of states is provided by the Wigner matrices (Peter-Weyl theorem) 
\be
\langle U| j,m,n\rangle= D^{(j)}_{mn}(U).
\ee
Equivalently, the state space ${\cal H}$ associated to the link decomposes as
\be
   {\cal H}=\oplus_j({\cal H}_j^* \otimes {\cal H}_j)
   \label{tensorproductstructure}
\ee
where ${\cal H}_j$ is the spin $j$ irreducible representation of $SU(2)$. The two factors are associated to the two ends of the links, respectively, as each transforms under the gauge transformation at one end.  The dynamics of the theory is obtained mapping these states to unitary representations of $SL(2,\mathbb{C})$. A unitary representation (in the principal series) \cite{Ruhl:1970fk} is labelled by a discrete spin $k\in\frac{1}{2}\mathbb{N}$ and a continuous parameter $p\in\mathbb{R}^+$ and the representation space is denoted ${\cal{H}}_{(p,k)}$. This space decomposes into irreducible representations of the $SU(2)$ subgroup as follows
\be
{\cal{H}}_{(p,k)}=\oplus_{j=k}^{\infty}{\cal{H}}_j
\ee
where ${\cal{H}}_j$ is the (finite dimensional) $SU(2)$ representation of spin $j$.  Therefore ${\cal{H}}_{(p,k)}$ admits a basis $|(p,k);j,m\rangle$ obtained diagonalizing the total angular momentum $L^2$ and the $L_z=\vec{L}\cdot \vec{z}$ component of the $SU(2)$ subgroup. The map that gives this injection, and defines the loop quantum gravity covariant dynamics is given by 
\bes
Y_\gamma: {\cal H}_{j} &\to&{{\cal H}^\gamma_j} \\ \nonumber
|j,m \rangle &\mapsto& |(\gamma j,j); j,m\rangle.  \nonumber
\ees 
 Here $\gamma \in \mathbb{R}^+$ is the Immirzi parameter. 

The two spin $j$ representations of Eq.\,\eqref{tensorproductstructure} are each mapped in this way, therefore the spin $j$ component of the state space lives in a subspace of 
\be
({\cal{H}}^\gamma_j)_{\cal T}^* \otimes ({\cal{H}}^\gamma_j)_{\cal S} 
\label{st}
\ee
where we have indicated by ${\cal S} $ and ${\cal T}$ the source and target ends of the link, respectively. On the image of the map $Y_\gamma$, the boost generator $\vec K$ and the rotation generator $\vec L$ satisfy
\be
\langle\vec{K}\rangle= \gamma\langle \vec{L}\rangle \label{simple}
\ee
as matrix elements.

In the classical theory, equation \eqref{simple} is satisfied by the conjugate momentum of the Lorentz connection, which becomes the $SL(2,\mathbb{C})$ generator in the quantum theory (see Appendix \ref{B}). This relation directly parallels \eqref{fgp} because (from the canonical analysis of the action) the area-vector is related to $\vec L$ by $\vec A=8\pi\gamma G \hbar \vec L$ and $\vec K$ is related to the energy by \eqref{aK}. 

\begin{figure}[t]
\includegraphics[width=2.5 in]{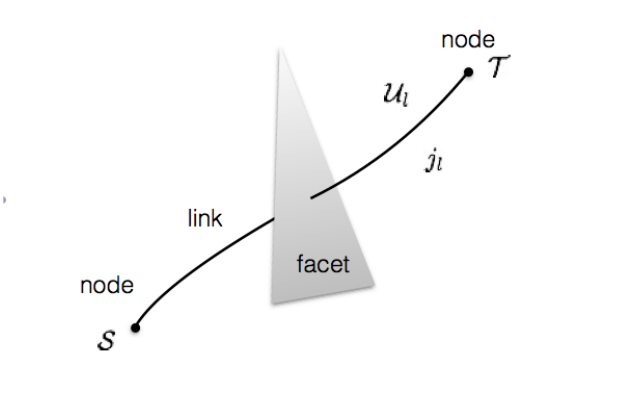}
\caption{Single link and its dual facet.}\label{link}
\end{figure}

The $SU(2)$ spin-network states represent the geometry of a spatial section of spacetime with normal $t^\mu$. If we measure the geometry of a spacelike surface with a certain precision, we can describe the result of our measurements by tessellating with $N$ facets $f$. Each facet is going to be described by a spin network link, which we can visualise as puncturing the surface \cite{Bianchi:2012ui}, see Fig.\,\ref{link}. Call $ |s_f\rangle$ the state describing a facet of a quantum surface of a given area and with normal in direction $\vec{n}$. This state has maximum magnetic number in direction $\vec{n}$. In writing it, for notational simplicity, one can restrict to $\vec{n}=\hat{z}$, then 
\be
|s_f\rangle = Y_{\gamma}|j,j\rangle = | (\gamma j, j);j, +j\rangle.
\ee 
The state $|s_f\rangle$ is an eigenstate of the area operator and of the the normal to the corresponding facet normalised to the area
\be
A_f^z |s_f \rangle=8\pi G \hbar \gamma {L}^z_f |s_f \rangle
=8\pi G \hbar \gamma j\, |s_f \rangle, \label{aa}
\ee
where $L_f^z=\vec{L}_f\cdot \hat{z}$. The states  $|s_f\rangle$ are mapped by $Y_{\gamma}$ to the unitary irreducible $SL(2,\mathbb{C})$ representation where the relation \eqref{simple} holds. For a single puncture, the relation $\langle\vec K_f \rangle=\gamma \langle\vec L_f\rangle$ then gives
\be
\langle K_f^z\rangle=\frac{\langle A_f^z\rangle}{8\pi G \hbar} \label{kk}
\ee
where $K_f^z$ is the boost dual to the facet surface. This relation gives immediately equation \eqref{fgp}, which is the basis of the second part of Jacobson's argument. From this relation we can obtain the Einstein equations.  Notice that this is a relation between the area of a space like surface (say in the $x$-$y$ plane) and the boost hamiltonian (in the dual $t$-$z$ plane).  As first observed by Smolin  \cite{Smolin:2012ys}, this is a direct way of deriving the Einstein equations from the covariant loop quantum dynamics. 

Next, observe that the tensor product structure in \eqref{st} is related to a local ``partition of space'' into two chunks, associated to the \emph{source} and \emph{target} nodes at the two ends of the link. In fact, the gravitational field operators, which are given by the left invariant and right invariant vector fields act respectively on the source and target end, and correspond to measurements performed on one side or the other of the facet \cite{Rovelli:2011eq}.   It therefore makes sense to define a half-link Hilbert space associated to each end of the link. The short-scale correlations of the theory are then captured by the correlations between these two Hilbert spaces, namely by the failure of the link states to be diagonal in this tensor decomposition. If  $|\Phi\rangle \in ({\cal{H}}^\gamma_j)^*\otimes({\cal{H}}^\gamma_j)$ is a pure state of the link, then its restriction to one half-link is the density matrix 
\be
\rho_{\scriptscriptstyle \cal T}=\tr_{{\cal{H}}_{\cal S }}[|\Phi\rangle \langle \Phi |], \label{linker}
\ee
that encodes the results of the measurements performed on the target side only of the facet (with an analogous relation holding for the other half of the link). The von Neumann entropy of this density matrix measures the entanglement across the facet. 

Let us now consider a particular family of states $|\Phi_0\rangle$ such that the associated reduced density matrix takes the the form 
\be
\rho_f=e^{-2\pi K_f}. \label{bbser}
\ee
where $K_f=\vec K_f\cdot \vec n$ is the boost generator in the direction normal to the facet (the notation $\rho_f=\rho_{\cal S}=\rho_{\cal T}$ indicates the symmetry of the reduced density matrix). We call these states ``Hadamard states" for a reason that will be clear below.  

These states have interesting properties. They have von Neumann entropy 
\be
S=-\tr[\rho_f \ln \rho_f]= 2\pi \,\tr[\rho_f K_f]= 2\pi \,\langle K_f \rangle,
\ee
where $\langle K_f \rangle= \langle \Phi_0 | K_f |\Phi_0 \rangle$ is the expectation value of $K_f$ on $| \Phi_0\rangle$. As shown in \eqref{dS}, a variation $\delta \rho$ relates  a change in the entropy to a change in the expectation value of $K_f$, namely
\be
\delta S= 2\pi\, \delta \langle K_f \rangle.
\ee
Therefore, given $\langle K_f \rangle=\gamma \langle L_f \rangle$, one finds
\be
\delta S= \frac{\langle\delta A_f\rangle}{4 G \hbar}, \label{abeke}
\ee
that is the area-entropy relation \eqref{beke}. We see here that the finiteness of the entropy density relies on the discreteness of the geometry at that scale and, in particular, on the finiteness of the spectrum of the area operator. 

Equation \eqref{fgp} is a classical relation while \eqref{kk} is associated to a quantum mechanical system. To clarify the relation between the two, we need to recall some basic structure in quantum theory. This always refers to results of a measurement performed by an external system.  Here the quantum system is a portion of quantum spacetime. The external observer is an accelerated observer which can measure local aspects of the geometry.  The states $|s_f\rangle$ describe a portion of a quantum surface, a small facet with given area $A_f$ and normal in direction $\vec{n}$. The evolution of the facet state in spacetime as seen by an accelerated observer is generated by the unitary operator representing a Lorentz boost
\be
|s_f^\tau\rangle=U(\tau)|s_f\rangle
\ee
where $U(\tau)=e^{\frac{i}{\hbar}H\,\tau}$, where the operator $H$ is 
\be
H=\hbar K_f\,a
\ee
with $a$ being the acceleration of the observer. This operator generates the evolution of the state along a specific isometry trajectory of the boost labelled by $a$. This trajectory represents the world line of an observer who moves in Minkowski space with uniform acceleration $a$. In this sense, a classical measurement of the facet state dynamics is associated to the covariant dynamics of the trajectory of the accelerated observer. This can be seen either passively as a motion of an observer in spacetime or actively as the evolution of spacetime seen by an observer.

Written in terms of the accelerated observer's quantities, the density matrix in \eqref{bbser} reads
\be
\rho_f=N\,e^{-\frac{2\pi}{\hbar a} H} \label{gibbser2}
\ee
and the expectation value of the Hamiltonian operator $H$ gives 
\be
E =\frac{A_f\,a}{8\pi G} \label{kk3}
\ee
where $E=\langle H\rangle$ has now the dimensions of an energy. This is exactly relation \eqref{fgp}, the core physical input from which the Einstein equations, as shown by Jacobson, follow in the classical limit.

To the \emph{ensemble} of single facet states given by \eqref{gibbser2}, the observer can effectively associate an absolute temperature, via the general definition
\be
T=\frac{\delta E}{\delta S}= \frac{a\hbar}{2\pi}.
\ee
because if it interacts with a large number of these, this is the temperature determining the transition probabilities between its eigenstates. This is the Unruh temperature.\footnote{Notice also that for a black hole the Hawking temperature at infinity is just the red shifted version of the near horizon temperature, and the asymptotic ADM energy is just the red shifted local energy. Therefore the standard black hole thermodynamics follows. } Therefore all the ingredients for Jacobson's derivation follow (see also \cite{Smolin:2012ys}).   

We have illustrated how geometrical discreteness arises in the formalism of loop gravity and shown that this discreteness gives rise to all of Jacobson's inputs microscopically. This line of thinking has exposed the possibility that local Lorentz invariance is encoded directly in the quantum correlations at the Planck scale and led us to introduce quantum gravitational Hadamard states. We further develop this idea below. 

\section{The architecture of semiclassical spacetime}\label{sec:architecture}

In  covariant loop quantum gravity \cite{Rovelli:2011eq} the dynamics is captured by a relation between the area of a spacelike surface (say in the $x$-$y$ plane) and the boost (in the $t$-$z$ plane). This is beautifully made explicit by the result of Jacobson, which shows that \eqref{fgp} yields the Einstein equations. 

The dynamical input that gives the Einstein equations in Jacobson's derivation is not the thermodynamics.  It is the relation between the boost generator and the area. This is the input that contains the Newton constant $G$ and gives the dynamics. Then, the presence of the horizon leads to the entanglement that gives rise to the entropy, and the relation between this entropy and the boost generator determines the KMS temperature.

In the case of a black hole, as shown in \cite{Frodden:2011eb}, the entropy at infinity, the energy at infinity (ADM mass) and the temperature at infinity (the Hawking temperature) are simply obtained by red shifting the local quantities to infinity. Their relation is precisely the local Clausius relation redshifted to infinity. This is unproblematic standard physics  (much like the Tolman law) and is well understood. 

These considerations do not put into question the tight constraints of reciprocal consistency that gravitational dynamics and thermodynamics of the vacuum state impose on each other; a consistency that Jacobson's derivation sheds light on. It is quite remarkable, indeed, that so much of the gravitational dynamics is constrained by consistency with the thermodynamics of quantum fields.  But thermodynamics does \emph{not} and \emph{cannot} give the full gravitational equations, as is clear for instance from the fact that gravitational theories different from general relativity are also compatible with the field thermodynamics \cite{Jacobson:2012yt}. 

What is the physical meaning of the Hadamard-like states introduced in Sec. \ref{sec:quantumgravity}?  These quantum states determine all the relations necessary for Jacobson's calculation. They are analogous to those obtained restricting the Minkowsky vacuum to the Rindler wedge. Like those states, they precisely describe the short scale correlations and the effect of local Lorentz invariance and the positivity of the energy.  Formally, a pure state on the link that gives rise to the density matrix \eqref{bbser} is simply built by having the source component of the state related to the target one by a Lorentz transformation with imaginary parameter $i\pi$
\be
\langle \Phi {| j,\vec{n}\rangle}_{\cal S}=e^{-\pi \gamma j}\, {\langle j,\vec{n}\,|}_{\cal T}.
\ee
In turn, this form of the state is formally determined by requiring a global Lorentz invariance and positivity of energy that allows for an analytic continuation \cite{Bisognano:1976za}.  Thus, these states can be viewed as implementing the local properties that render a semiclassical geometry locally Lorentz invariant.  If every link of a spin network carries the correlations of these states, then the corresponding quantum state has everywhere the short scale correlations of a conventional quantum field theory state. These are therefore the states that may build up the architecture of a classical spacetime geometry \cite{Bianchi:2012ev}.  In quantum field theory on curved spacetime attention is restricted to states having an appropriate short scale behavior, which is determined by it being the same as that of a flat Lorentz invariant quantum field theory. States with this local property are called Hadamard states. This is why, by analogy, we call the states of Sec.\,\ref{sec:quantumgravity} Hadamard states: they behave like quantum field states in the small. Notice that the relation here is not at the infinitesimal level, but at the finite Planck-scale, namely across each single link of a spin network. Quantum gravity is a theory without infinities.

\section{Conclusion}

Jacobson's derivation of the Einstein equations  admits an interpretation different from the common one where it is assumed to indicate the existence of microscopic states for which the Einstein equations would be an equation of state. The alternative interpretation is based on the identification of the relevant dynamics simply as the quantum version of Einstein's dynamics.

The relevant microscopic degrees of freedom are the quanta of the gravitational field. These are discrete at the Planck scale, and this is the source of the finiteness of the entropy. The interaction of any system with the gravitational field in a region of spacetime is described by the field operators of the quantum theory. As for the electromagnetic field, some of the quantum states of the gravitational field have a semiclassical behavior. These appear as continuous geometries when probed at large scales: the mean value of the field operators determines the value of the geometrical observables. The approximation holds only in the semiclassical limit, namely at scales larger than the Planck scale. These states present short scale correlations, like all the finite energy states of conventional quantum field theory. But these correlations do not involve modes of arbitrarily high frequency, because of the Planck scale quantum discreteness. 

Because of the universality of the gravitational coupling, and thanks to a mechanism that can be intuitively understood as explained in Section \ref{eu}, the entropy density measuring these correlations is universal. An external observer accelerating in the semiclassical geometry interacts only with observables on one side of its causal horizon and can therefore associate an entropy density to the horizon, which is finite and universal. The dynamics ties any change in the area with an energy flow, therefore, because of the universality, the change in the area is tied by the dynamics to a change in entropy, with a universal temperature $T$. This temperature is determined by the entropy-energy relation and therefore, ultimately, simply by the local Lorentz invariance and the positivity of the energy of the local states showing short scale correlations. In spite of the fact that the source of the entropy is entanglement, this temperature behaves as a true temperature because the density matrix is a KMS state with respect to the flow that describes the evolution of an observer accelerating in the mean geometry, and because this observer interacts with a large number of degrees of freedom, all in this KMS state. It follows from this that $T$ is detected by a thermometer moving with the observer. 

The relation \eqref{fgp} between area and energy is determined by the quantum dynamics of general relativity: it is true both in the quantum and the classical theory. The universal relation between area and entropy is determined by the short scale structure of the semiclassical states, their local Lorentz invariance and the positivity of the energy, as in the Bisognano-Wichmann theorem. In the deep ultraviolet this picture is confirmed by loop quantum gravity, where all the input formulas of Jacobson's derivation can be recovered starting from the elementary quantum dynamics of the theory, for appropriate states that have semiclassical properties, which we have called Hadamard states.

The entropy in Jacobson's calculation is therefore entanglement entropy across the horizon. It is a property of a pure state, when restricted to the sub algebra of observables lying on one side of the causal horizon. It is not related to any mysterious underlying degrees of freedom. The entropy balance relation does not imply statistical uncertainty and in general holds also for entanglement entropy. 

What Jacobson has shown is (i) that the area-energy relation \eqref{fgp}, combined with general covariance, determines the Einstein equations, a result in line with similar previous derivations of the Einstein equations from a single local equation, e.g. \cite{Baez:2001qy}; and (ii) that this dynamics not only implies the universality of the entropy/area ratio, but is also implied by it.

This does not diminish the relevance of Jacobson's result. To the contrary, it puts it in an even more interesting light. Jacobson's remarkable discovery is not that there are mysterious microstates beyond the quantized gravitational field. It is that the dynamics of gravity and the Einstein equations can be recovered from a simple relation between boost generator and area of its dual surface, and these quantities are related to the entanglement entropy in semiclassical states. This is a step ahead in unraveling the spectacular beauty and the simplicity of general relativity, not an indication of its limits.

\centerline{---------}

Thanks to Eugenio Bianchi for crucial inputs in the species problem and for numerous discussions. HMH acknowledges support from the National Science Foundation (NSF) International Research Fellowship Program (IRFP) under Grant No. OISE-1159218.

\vspace{4em}

\appendix 

\section{KMS and thermometers}\label{A}

Consider a quantum system described by an observable algebra $\cal A$, with observables $A,B,...$, which is in a state $\sigma:{\cal A} \to {\mathbb C}$.  Let $\alpha_t:{\cal A} \to {\cal A}$ with $t\in  {\mathbb R}$ be a flow on the algebra.  We say that the state $\sigma$ is an equilibrium state with respect to $\alpha_t$ if for all $t$, 
\be
\sigma(\alpha_tA)=\sigma(A),   \label{uno}
\ee
and that its temperature is $T=1/\beta$ if 
\be
f_{AB}(t)=f_{BA}(-t+i\beta)   \label{due}
\ee
where
\be
f_{AB}(t)=\sigma(\alpha_tA\, B).  \label{f}
\ee
An equivalent form of \eqref{due} is given in term of the Fourier transform $\tilde f_{AB}$ of $f_{AB}$ as
\be
\tilde f_{AB}(\omega)=e^{-\beta \omega }\tilde f_{AB}(-\omega).
  \label{duedue}
\ee

The canonical example is provided by the case where $\cal A$ is realized by operators on a Hilbert space $\cal H$ where a Hamiltonian operator $H$ with eigenstates $|n\rangle$ and eigenvalues $E_n$ is defined, the flow and the state are defined in terms the evolution operator $U(t)=e^{-iHt}$ and the density matrix $\rho=e^{-\beta H}=\sum_ne^{-\beta E_n}|n\rangle\langle n|$, by 
\be
   \alpha_t(A)=U^{-1}(t)AU(t)
\ee
and 
\be
   \sigma(A)={\rm tr}[\rho A]
\ee
respectively. In this case, it is straightforward to verify that equation \eqref{uno} and \eqref{due} follow.  

However, the scope of equations \eqref{uno} and \eqref{due} is wider than this canonical case. For instance, these equations permit the treatment of thermal quantum field theory, where the Hamiltonian is ill-defined, because of the infinite energy of a thermal state in an infinite space.  It is indeed important to remark that these two equations capture immediately the physical notions of equilibrium and temperature.  For equation \eqref{uno}, this is pretty obvious: the state is in equilibrium with respect to a flow of time if the expectation value of any observable is time independent.  

The direct physical interpretation of equation \eqref{due} is less evident: it describes the coupling of the system with a thermometer.  

To see this, consider a simple thermometer formed by a two-state system with an energy gap $\epsilon$, coupled to a quantum system $S$ by the interaction term
\be
   V=g(|0\rangle\langle 1|+|1\rangle\langle 0|) A. \label{V}
\ee
where $g$ is a small coupling constant. The amplitude for the thermometer to jump up from the initial state $|0\rangle$ to the final state $|1\rangle$, while the system moves from an initial state $|i\rangle$ to a final state $|f\rangle$ can be computed using Fermi's golden rule to first order in $g$:
\begin{eqnarray}
     W_+(t)&=&g \int_{-\infty}^t  dt\ (\langle 1|\otimes \langle f|)\ \alpha_t(V)\  (|0\rangle+| i \rangle)  \nonumber \\
&=&g \int_{-\infty}^t  dt \ e^{it\epsilon }\langle f|\alpha_t(A) | i \rangle.
\end{eqnarray}
The probability for the thermometer to jump up is the modulus square of the amplitude, summed over the final state. This is 
\begin{eqnarray}
     P_+(t)&=&g^2 \int_{-\infty}^t  dt_1 \int_{-\infty}^t  dt_2 \ e^{i\epsilon(t_1-t_2) } \sigma(\alpha_{t_2}(A^\dagger)\alpha_{t_1}(A)) \nonumber
\end{eqnarray}
where we have used the algebraic notation $\omega(A)= \langle i|A| i \rangle$. If the initial state is an equilibrium state 
\begin{eqnarray}
     P_+(t)&=&g^2 \int_{-\infty}^t  dt_1 \int_{-\infty}^t  dt_2 \ e^{i\epsilon(t_1-t_2) } f_{A^\dagger\!A}(t_1-t_2).\nonumber
\end{eqnarray}
and the integrand depends only on the difference of the times. The transition probability per unit time is then 
\begin{eqnarray}
     p_+&=& \frac{dP_+}{dt} = g^2\  \tilde f_{A^\dagger\!A}(\epsilon)
\end{eqnarray}
which shows that \eqref{f} is precisely the quantity giving the transition rate for a thermometer coupled to the system. 
It is immediate to repeat the calculation for the probability to jump down, which gives 
\begin{eqnarray}
     p_-&=&  g^2\ f_{AA^\dagger}(-\epsilon).  \label{V22}
\end{eqnarray}
And therefore \eqref{duedue} expresses precisely  the fact that the thermometer thermalizes at temperature $\/\beta$, that is $p_+/p_-=e^{-\beta \epsilon}$

These observations show that equations \eqref{uno} and \eqref{due} define a generalization of the standard quantum statistical mechanics, which fully captures thermal properties of a quantum  system.  The structure defined by these equations is called a modular flow in the mathematical literature: $\alpha_t$ is a modular flow, or a Tomita flow, for the state $\omega$, and is the basic tool for the classification of the $C^*$ algebras. In the physical literature, the state $\omega$ is called a KMS state (for the time flow).  In order to show that a system is in equilibrium and behaves thermally in a certain state, it is sufficient to show that the state satisfies these two equations. In order for a KMS state to behave as an equilibrium state at some temperature, however,  the corresponding flow must be the evolution flow of the thermometer.  Therefore the claim here is not that any KMS state, with respect to any flow, behaves as a thermal state. 
 
Notice that in the context of Hilbert space quantum mechanics, a state can satisfy these two equations also if it is a pure state. The standard example is provided by the vacuum state of a Poincar\'e invariant quantum field theory, which is KMS with respect to the modular flow defined by the boost in a given direction. Being Poincar\'e invariant, the vacuum is invariant under this flow, and a celebrated calculation by Unruh shows that it is KMS at inverse temperature $2\pi$.  

In this case, the physical interpretation is simply given by the fact that an observer stationary with respect this flow, namely an accelerated observer at unit acceleration, will measure a temperature $1/2\pi$. Unruh's original calculation, indeed, follows precisely the steps above in equations (\ref{V}-\ref{V22}). 

\section{The proportionality between area and boost generator}\label{B}

The Einstein equations can be derived from the Holst action  \cite{Hojman:1980kv,Holst:1996fk}
   \be\label{Palatini}
S[e,\omega]= \frac1{8\pi G} \left(\int (\star e\wedge e)\wedge F+\frac1{\gamma} \int e\wedge e\wedge F\right)
~.
   \ee
where $e$ is the (co)tetrad one-form, $F$ the curvature of the Lorentz connection $\omega$, the star indicates the hodge dual on internal indices and a contraction on the internal indices is implicit.  On a  spacelike boundary $\Sigma$, the quantity 
\be
\Pi= \frac1{8\pi G} (\star e\wedge e+ \frac{1}{\gamma} e\wedge e)
\label{Bmomentum}
\ee
is the derivative of the action with respect to $\partial\omega/\partial t$, and therefore is the momentum conjugate to the connection.  The (co)tetrad $e$ can be chosen to (locally) map $\Sigma$ into a spacelike 3d linear subspace of Minkowski space. 
 The subgroup of the Lorentz group that leaves this subspace invariant is the $SO(3)$ {rotations} subgroup, and its existence breaks the local $SO(3,1)$ invariance down to $SO(3)$ at the boundary. That is, the boundary allows us to pick up a preferred Lorentz frame. 

In coordinates, we can see this by writing the unit length vector $n_I$ normal to all vectors tangents to $\Sigma$, that is
\be 
n_I\sim  \epsilon_{IJKL}\ e_\mu^J\, e_\nu^K\, e_\rho^L\ \frac{\partial x^\mu}{\partial \sigma^1}\frac{\partial x^\nu}{\partial \sigma^2}\frac{\partial x^\rho}{\partial \sigma^3}
\ee
where $x^\mu(\sigma)$ is the imbedding of the boundary $\Sigma$ into spacetime. We can then use $n_I$  to gauge fix $SO(3,1)$ down to the $SO(3)$ subgroup that preserves it. We can orient our local Lorentz frame in such a way that the boundary is locally a fixed-time surface, so that $n_I=(1,0,0,0)$.  

The pull back to $\Sigma$  of the momentum two-form $\Pi$ can be decomposed into its electric  $K^I=n_J\Pi^{IJ}$ and magnetic  $L^I=n_J(\star \Pi)^{IJ}$ parts, in the same manner in which the electromagnetic tensor $F^{IJ}$ can be decomposed into electric and magnetic parts once a Lorentz frame is chosen. Since $\Pi$ is antisymmetric, 
$L^I$ and $K^I$ do not have components normal to $\Sigma$, that is $n_IK^I=n_IL^I=0$ and are therefore three dimensional vectors in the space normal to $n$, which we can denote $\vec K$ and $\vec L$. In the gauge $n_I=(n,0,0,0)$ these are simply 
\be
           K^i=\Pi^{i0}, \hspace{3em}     L^i=\frac12 \epsilon^i{}_{jk} \Pi^{jk}. 
\ee
where we write $\vec K=\{K^i\}$ and $\vec L=\{L^i\}$, with  $i=1,2,3$, which are relations analogous to the ones that relate electric and magnetic fields to the Maxwell tensor.  From the definitions, we have 
\be
n_I e^I_\mu|_\Sigma = 0 \; ,
\ee
therefore at the boundary
\be
n_I (\star\Pi)^{IJ} = \frac{1}{\gamma}  n_I \Pi^{IJ} \;,
\ee
which also reads
\be
\vec K=\gamma \vec L \, .
\label{KgL}
\ee 
This is called the linear simplicity relation  \cite{Engle:2007wy}.  Decompose $\vec K=K\vec n$ and, because
\be
L^i = \frac{1}{8\pi G\gamma} n_J (e\wedge e)^{iJ}
\ee
it follows
\be
         \vec L=\frac{1}{8\pi G \gamma} A\vec n,
\ee 
where $A$ is the area element for a surface element normal to the direction $\vec n$ of the boost. Therefore
\be
          K=\frac{A}{8\pi G}.
\ee 
The boost generator $K$ is related to the area element.  An observer with acceleration $a$ moves along a flow generated by $H=aK$, therefore his energy is 
\be
    E=a K=\frac{aA}{8\pi G}.
\ee
which is the Frodden-Gosh-Perez relation. 

\section{Jacobson derivation}\label{C}

We recall here the main steps of the second part of Jacobson's derivation: from equation \eqref{fgp} to the Einstein's equations. The change of the horizon area with the expansion of the null geodesics comprising it is 
\be
\delta A= \int_{H}\vS\,  \theta \,d\lambda \label{arria}\,, 
\ee
where $\vS$ is the 2-surface area element of the horizon cross-section, $\lambda$ s the affine parameter along the null geodesics and $\theta$ is the usual expansion.
For a small horizon perturbation, the infinitesimal evolution of $\theta$ is given by a linear expansion around its equilibrium value at $p$, up to the first order in $\lambda$,
\be
\theta \approx \theta_p+\lambda \, \left. \frac{d\theta}{d \lambda} \right|_p\,+\mathcal{O}(\lambda^2) \label{exp}\,. 
\ee
At equilibrium, the $0th$ order contribution to the area change $\theta_p$ vanishes. The first order contribution is given by the Raychaudhuri equation,
\be
\frac{d\theta}{d \lambda}= -\frac{1}{2}\theta^2-\|\sigma\|^2-R_{\mu \nu}\ell^{\mu}\ell^{\nu} \label{Ray}\, , 
\ee
where $\|\sigma\|^2$ stands for the squared congruence shear $\sigma^{\mu\nu}\sigma_{\mu\nu}$ and $\ell^\mu$ is the affine tangent vector to the congruence. 
If the small deformation is caused by some matter stress tensor at order $\epsilon$, then the metric perturbation and therefore $\theta$ and $\sigma$ will be of order $\epsilon$. Hence the term $(1/2\theta^2-\|\sigma\|^2)$ can be neglected within an equilibrium setting \cite{jacopa, eling07, chirco10}. Therefore, the entropy variation takes the form
\be 
\delta S=\alpha\delta A = - \alpha  \int_{H}  \vS \, \lambda\,(R_{\mu \nu}\ell^{\mu}\ell^{\nu})_p \,d\lambda \label{ee}\, .
\ee

The mean energy variation of the thermal system, given by the boosted energy current flux of matter, can be described in terms of heat, as it is perfectly thermalized by the horizon. Thereby, one can define the heat flux across the horizon as
\be 
\delta E=  \int_{H}  T_{\mu\nu}\chi^{\mu} d\Sigma^{\nu} \label{hea} \, ,
\ee
where $T_{\mu\nu}$ is the matter stress energy tensor, $\chi^{\mu}$ is the local horizon approximate Killing vector, while the volume element is given by $d\Sigma^\nu=\ell^\nu\, \vS \, d\lambda$. Given the local Killing character of the local Rindler horizon, equation \eqref{hea} can be expressed in terms of the null congruence vectors, 
\be 
\delta E=  \int_{H} \vS \, d\lambda \,\, (-\lambda \kappa)\,\,T_{\mu \nu}\ell^{\mu}\ell^{\nu}  \label{heaa}\, . 
\ee

At this point, asking for relation (\ref{clau}) to hold for all null vectors $\ell^\mu$, one can equate at $p$ the $\mathcal{O}(\lambda)$ integrands in (\ref{ee}) and (\ref{heaa}), obtaining
\be 
\frac{2\pi}{\hbar \alpha}\, T_{\mu \nu} \ell^{\mu} \ell^{\nu}= R_{\mu \nu}\ell^{\mu}\ell^{\nu} \label{eoss}\, 
\ee
and thereby, in any frame,
\be
\frac{2\pi}{\hbar \alpha}\,T_{\mu \nu}= R_{\mu \nu}\,+\,\Phi\,g_{\mu \nu} \label{eos0} \, ,
\ee
where $\Phi$ is an undetermined integration function. 

Eventually, by assuming the local energy conservation, that is $\nabla^{\nu} T_{\mu \nu}=0$, applying the divergence operator on both sides of (\ref{eos0}), and using the contracted Bianchi identity $\nabla^{\nu}R_{\mu \nu}=\frac{1}{2}\nabla_{\mu} R$, one finally gets $\Phi=-\frac{1}{2}R-\Lambda$, hence
\be
\frac{2\pi}{\hbar \alpha} \,T_{\mu \nu}= R_{\mu \nu}\,-\frac{1}{2}R\,g_{\mu \nu}\, -\,\Lambda\, g_{\mu \nu}\, , \label{eos1}
\ee
where $\Lambda$ is some arbitrary integration constant.
Once the condition
\be
\alpha=\frac{1}{4 \hbar G} \label{alfa}
\ee
is imposed, \eqref{eos1} gives the Einstein equations. The equivalence principle implies that the above construction can be done at {\em any} spacetime point $p$, hence equation \eqref{eos1} holds in any point of a pseudo-riemannian spacetime \cite{Jacobson:1995ab}.

\end{document}